\documentclass{article}
\usepackage{spconf,amsmath,graphicx}
\usepackage[hidelinks]{hyperref}
\usepackage{booktabs}

\title{Towards Deployable Underwater Vessel Classification}

\name{Abishek Soti,
Thura Pyae Sone,
Naqib Ibnul,
Htoo Htet Aung,
Henry Zhong,
Gregory Cohen,
Ying Xu}
\address{ICNS, Western Sydney University, Sydney, Australia}

\begin{document}

\ninept
\maketitle
\begin{abstract}
We propose a compact underwater acoustic classification framework combining multi-representation feature engineering, temporal statistical pooling, and compact convolutional architectures designed for acoustic time–frequency and cochlear representations. We investigate multiple conventional and auditory-inspired representations and first evaluate lightweight classifiers and Conventional Neural Networks (CNNs) on ShipsEar dataset. On the provided split, a two-layer CNN achieves a macro F1 of 0.9918, while a Radial Basis Function Support Vector Machine (RBF-SVM) reaches 0.9883. However, source-recording provenance cannot be reconstructed, preventing verification of recording-independent generalisation. We therefore evaluate on DeepShip dataset using recording-level partitioning before segmentation. Under this protocol, a 157K-parameter compact CNN achieves a test macro F1 of 0.7226, while an 11.17M-parameter ResNet18 provides no improvement in validation performance under the matched setting. These results demonstrate the importance of representation-aware feature and model design, together with rigorous recording-level evaluation, for classification performance and deployability in compact underwater acoustic systems.
\end{abstract}




\section{Introduction}
\label{sec:intro}


Autonomous passive-acoustic platforms such as gliders and profilers
operate under tighter power, computation, and communication constraints,
making compact embedded processing desirable \cite{toma2018smart}.
For this deployment setting, we explore engineering the audio time-frequency representations relying on the characteristics of underwater vessel acoustics to improve classification performance instead of increasing model capacity.


Building on this, audio feature representations becomes one of our central design consideration. Different time-frequency representations capture different spectral and temporal characteristics of underwater vessel acoustics, and we will explore if these differences can affect their classification performance. We therefore consider conventional representations including the short-time Fourier transform (STFT), Mel spectrograms, mel-frequency cepstral coefficients (MFCC), and constant-Q transform (CQT) \cite{brown1991cqt}, alongside auditory-inspired approaches such as gammatone filterbanks \cite{hohmann2002gammatone} and the Cascade of Asymmetric Resonators with Fast-Acting Compression (CARFAC) \cite{lyon2017human,lyon2024carfacv2} to assess whether these auditory-inspired representations can improve classification performance without requiring greater model complexity. Similarly, Low-Frequency Analysis and Recording (LOFAR) spectrum restricted to $0$--$3$~kHz   was evaluated on ShipsEar, a public underwater acoustic dataset containing recordings from multiple vessel types commonly grouped into five classes \cite{santosdominguez2016shipsear}, and reported $98.50\%$ accuracy using a $21.61$M-parameter Mobile\_ViT \cite{yao2024mobilevit}. In comparison, our STFT two-layer CNN achieves $99.10\%$ accuracy with approximately $183$K parameters on the available standard ShipsEar split. Beyond this compact CNN result, the representations are also evaluated using classical classifiers and shallow Transformer architectures to examine how model choice affects performance without requiring excessive model capacity.


Concerned by the near-ceiling performance ($99.10\%$ accuracy) achieved by our STFT two-layer CNN on the provided ShipsEar split, we examine whether the evaluation protocol may contribute to the observed performance. The processed ShipsEar data do not retain sufficient source-recording provenance to verify that segments from the same original recording are confined to only one of the training, validation, or test partitions. If acoustically similar segments from the same recording occur across partitions, measured performance may overestimate generalisation to unseen recordings. We therefore extend the evaluation to DeepShip, a larger underwater vessel dataset containing four classes (Cargo, Passenger, Tanker, and Tug), in which source recordings can be identified and partitioned
before segmentation \cite{irfan2021deepship}. The original DeepShip study reported its strongest result of $77.53\%$ accuracy using CQT features with a separable convolutional autoencoder \cite{irfan2021deepship}.

Using recording-level separation on DeepShip, we test whether the strong
compact-model performance observed on the provided ShipsEar split persists
when evaluation is restricted to unseen source recordings. Conventional and
auditory-inspired feature representations are evaluated using classical
classifiers and compact neural models, while ResNet18
\cite{He_2016_CVPR} provides a substantially higher-capacity reference.
Under this setting, the compact CNN achieves higher validation macro F1 than
ResNet18 despite using approximately $71\times$ fewer parameters, while
performance varies substantially across acoustic representations. ShipsEar therefore serves as an initial diagnostic benchmark, whereas DeepShip provides the primary recording-level evaluation and shows that increasing model capacity alone does not improve generalisation. Additionally, to examine which spectral regions contribute to the compact CNN's predictions, we apply Grad-CAM \cite{selvaraju2017gradcam}, revealing class-dependent frequency
relevance.


\section{Datasets and Evaluation Protocol}
\label{sec:data}

\subsection{ShipsEar and DeepShip}

ShipsEar contains approximately $3.1$ hours of audio from $90$ recording
sessions covering $11$ vessel types sampled at $64$ kHz and then
resampled to $16$ kHz, commonly grouped into five classes
\cite{santosdominguez2016shipsear}. DeepShip contains recordings from
$265$ vessels across four classes---Cargo, Passenger, Tanker, and Tug
\cite{irfan2021deepship}. It contains $609$ readable recordings
comprising approximately $47.2$ hours of audio.

\subsection{Partitioning and Evaluation}

The processed ShipsEar distribution provides a fixed train--test split of $1778$ and $445$ pre-segmented $5$ second clips. Because sufficient source-recording identifiers are unavailable for the Shipsear dataset, we cannot verify that clips originating from the same recording are confined to only one partition.  DeepShip recordings are assigned to training, validation, and test partitions before segmentation so that no source recording contributes segments
to more than one partition. The main representation experiments use
non-overlapping 5-second segments resampled to 16 kHz, resulting in
26,933 training, 3,266 validation, and 3,457 test segments. The separate ResNet18 reproduction follows the published $3$ second segmentation and $32$ kHz sampling protocol.
For the main experiments using recording-level partitioning, accuracy and macro F1-score are reported, with macro F1 used as the primary metric. Validation macro F1 is used for model selection and early stopping, while the test partition is reserved for final evaluation.


\section{Representations and Models}
\label{sec:methods}

\begin{figure}[h]
    \centering
    \includegraphics[width=\columnwidth]{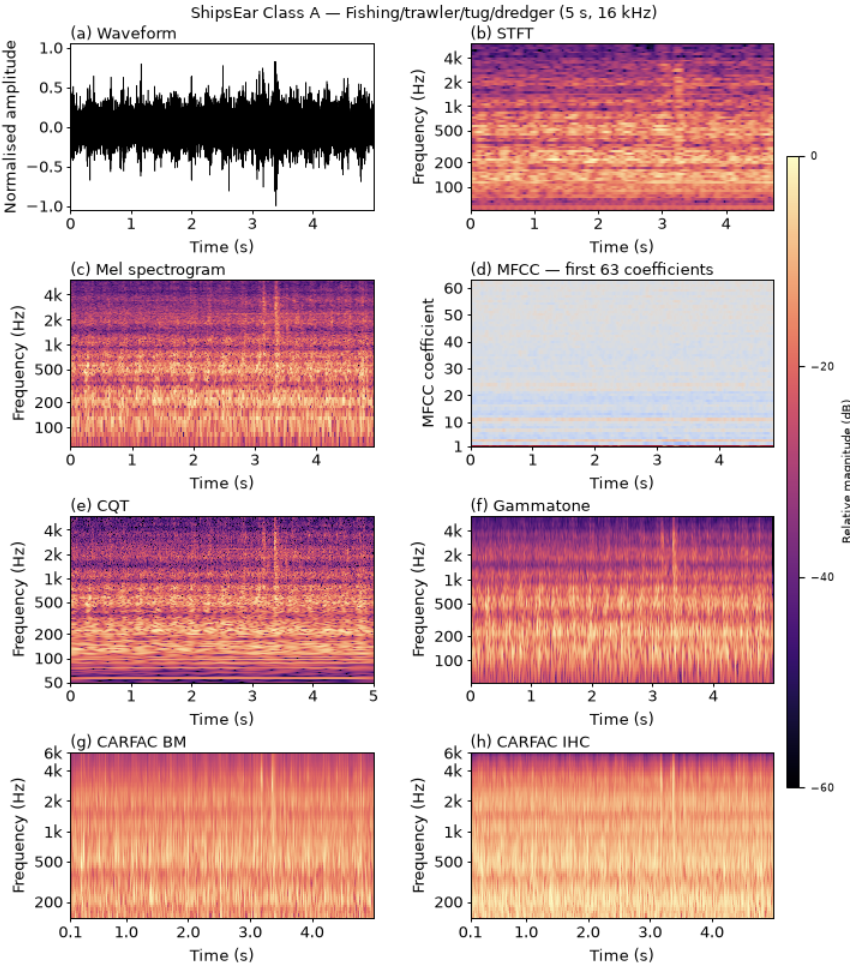}
    \caption{Example ShipsEar waveform and corresponding conventional
    and auditory-inspired acoustic representations used in this study.}
    \label{fig:representations}
\end{figure}

\subsection{Acoustic Representations and Statistical Pooling}

For each audio segment, we generate STFT, Mel, MFCC, CQT, Gammatone,
and CARFAC representations. The CARFAC outputs considered are the
basilar-membrane (BM) response, inner-hair-cell (IHC) response, and
lateral-inhibition (LI) output. Figure~\ref{fig:representations} shows
the same ShipsEar audio example across these representations. Figure \ref{fig:representations} shows the same ShipsEar audio
example across these representations.

For the classical classifiers, each time--frequency representation is
compressed along the temporal dimension by computing the mean and
standard deviation of each frequency or filter channel across time.
The resulting statistics are concatenated to form a fixed-length feature
vector that is provided to logistic regression or an RBF-SVM. For
example, a 128-bin STFT representation is reduced to 256 values,
consisting of one mean and one standard deviation for each frequency
bin. Similar statistical pooling is applied to learned CNN feature maps
before classification, with the specific pooling arrangement for the
deeper DeepShip model described in Section~3.2.

To examine the effect of statistical pooling independently of neural
model configuration, we compare the ShipsEar STFT RBF-SVM with and
without pooling. Table~\ref{tab:pooling_ablation} shows that
mean--standard-deviation pooling reduces the input dimensionality from
$60{,}800$ to $256$ while improving test macro F1 from $0.9641$ to
$0.9883$.

\begin{table}[h]
\centering
\caption{Effect of temporal statistical pooling on ShipsEar STFT classification.}
\label{tab:pooling_ablation}
\setlength{\tabcolsep}{4pt}
\begin{tabular*}{\linewidth}{@{\extracolsep{\fill}}llrrr@{}}
\toprule
\textbf{Model} &
\textbf{Pooling} &
\textbf{Input dim.} &
\textbf{Params} &
\textbf{Test F1} \\
\midrule
RBF-SVM & None & 60,800 & -- & 0.9641 \\
RBF-SVM & Mean+std & 256 & -- & \textbf{0.9883} \\
2-layer CNN & None & 302,080 & 1.668M & 0.8987 \\
2-layer CNN & Mean+std & 5,120 & 183K & \textbf{0.9698} \\
\bottomrule
\end{tabular*}
\end{table}

The scikit-learn classifiers used in the Shipsear dataset from table~\ref{tab:pooling_ablation} were also deployed on a Raspberry Pi with a lightweight interface for audio-file selection and prediction, providing a proof-of-concept implementation for embedded inference.

\subsection{Neural Models and Model Capacity}

The principal compact convolutional model is a four-layer TinyVGG-style
CNN with $32$, $32$, $64$, and $64$ channels and kernels of
$5\times15$, $9\times9$, $3\times3$, and $3\times3$. Spatial
$2\times2$ max pooling follows the second and fourth convolutional
stages. Statistical pooling is then applied to the learned feature maps
before classification, reducing the dimensionality of the final
representation while retaining summary information from the learned
responses. Models are trained using AdamW and class-weighted
cross-entropy, with validation macro F1 used for model selection and
early stopping.

For the ShipsEar experiments, we also evaluate a shallow Patch
Transformer that operates directly on the two-dimensional acoustic
representations. Each representation is divided into local patches,
which are projected into token embeddings and processed by a lightweight
Transformer encoder before classification. This provides a
non-convolutional comparison while retaining substantially lower model
capacity than large pretrained audio Transformers. 

Additionally, a  standard ResNet18 \cite{He_2016_CVPR} provides the higher-capacity
convolutional comparison, with $11.17$M parameters compared with
approximately $157$K for the compact DeepShip CNN. It is evaluated under
our $5$~s recording-level DeepShip protocol and separately in a
protocol-close reproduction of Chen et al. \cite{chen2026uatfsn},
following their reported $3$ second, $32$ kHz, $80/20$ recording-level split,
128-bin Mel representation, 2048-point FFT, and 80-epoch configuration.
Methodological choices not specified in the publication are treated as
reproduction assumptions.

\subsection{Model Explanation}

Grad-CAM \cite{selvaraju2017gradcam} is applied to the final
convolutional layer of the four-layer DeepShip STFT CNN to examine which
spectral regions contribute to its vessel-class predictions. Grad-CAM
maps are generated from correct predictions on the DeepShip test set and
normalized within each sample. The resulting maps are aggregated by
vessel class to obtain class-wise frequency-relevance profiles.

\subsection{ShipsEar as a Diagnostic Benchmark}

ShipsEar provides a useful diagnostic of how strongly measured
classification performance can depend on the available evaluation
setting. Table~\ref{tab:shipsear_results} summarises selected in-house
results obtained from the provided fixed $5$ second split together with
selected published results for context. The table reports representative
configurations rather than the complete classifier--feature sweep.

\begin{table}[h]
\centering
\setlength{\tabcolsep}{1.5pt}
\renewcommand{\arraystretch}{1.0}

\caption{Selected ShipsEar classification results under the available
fixed $5$~second split.}
\label{tab:shipsear_results}

\begin{tabular*}{\columnwidth}{@{\extracolsep{\fill}}p{0.19\columnwidth}p{0.24\columnwidth}ccc@{}}
\toprule
\textbf{Feature} &
\textbf{Model} &
\textbf{Params} &
\textbf{Acc.} &
\textbf{Macro F1} \\
\midrule

STFT & 2L-CNN & 183,237 & \textbf{0.9910} & \textbf{0.9918} \\
STFT & RBF-SVM & N/A & 0.9888 & 0.9883 \\
CQT & 2L-CNN & 208,037 & 0.9820 & 0.9839 \\
Mel & 2L-CNN & 208,837 & 0.9775 & 0.9789 \\
LI & RBF-SVM & N/A & 0.9707 & 0.9712 \\
MFCC & RBF-SVM & N/A & 0.9663 & 0.9677 \\
IHC & 2L-CNN & 177,637 & 0.9617 & 0.9640 \\
BM & 2L-CNN & 199,525 & 0.9595 & 0.9606 \\

\midrule

STFT & Patch Transformer & 303,621 & 0.9708 & 0.9711 \\

\midrule

LOFAR (STFT, $0$--$3$ kHz)
& Mobile\_ViT \cite{yao2024mobilevit}
& 21.61M
& 0.9850
& 0.9838 \\

ZCR, RMS, MFCC,Chroma
& DCMT \cite{mahmud2026dcmt}
& 0.70M
& 0.9819
& -- \\

Mel(128 bin)
& ResNet18 \cite{yan2026multhead}
& 11.2M
& 0.6396
& -- \\

Mel(128 bin)
& Seq. multi-head \cite{yan2026multhead}
& 11.2M
& 0.7031
& -- \\

\bottomrule
\end{tabular*}

\vspace{1mm}

\noindent
Rows above the final divider are from this study, rows below are taken from the cited publication and may use different evaluation protocols. "--" indicates an unreported metric.

\end{table}

Among our experiments, the two-layer STFT CNN achieves $0.9910$
accuracy and $0.9918$ macro F1, while the pooled STFT RBF-SVM reaches
$0.9888$ accuracy and $0.9883$ macro F1. The difference in development
cost is substantial: on the constrained $8$~GB development system used
for these experiments, the RBF-SVM trains in less than two minutes,
whereas the corresponding ShipsEar CNN experiments require approximately
$4.5$ hours. This makes the classical model particularly useful for rapid
iteration while retaining nearly the same classification performance.

The shallow Patch Transformer provides a non-convolutional comparison.
Its strongest configuration uses STFT with $303{,}621$ parameters and
reaches $0.9708$ accuracy and $0.9711$ macro F1. This remains below both
the two-layer STFT CNN and the pooled STFT RBF-SVM, showing that greater
architectural complexity does not improve performance under this
ShipsEar evaluation.

The final four rows of Table~\ref{tab:shipsear_results} are taken from
published studies rather than reproduced by us. Yao et al.
\cite{yao2024mobilevit} report $0.9850$ accuracy and $0.9838$ macro F1
using a Mobile\_ViT model with a LOFAR/STFT representation restricted to
$0$--$3$~kHz, while Mahmud et al. \cite{mahmud2026dcmt} report $0.9819$
accuracy using their DCMT model with ZCR, RMS, MFCC, and Chroma features.
The recording-level results of Yan et al. \cite{yan2026multhead} are
included separately because their evaluation protocol differs
substantially from the provided ShipsEar split used for our experiments.

The near-ceiling in-house scores should therefore be interpreted
cautiously because the source-recording identities of the processed
ShipsEar clips cannot be reconstructed sufficiently to verify
recording-level independence. The historical exploratory sweeps also
used the provided evaluation split during configuration and epoch
selection, so these results are retained as diagnostic benchmarks rather
than unbiased estimates of performance on unseen recordings. Yan et al.
\cite{yan2026multhead}, who explicitly separated ShipsEar recordings,
obtained substantially lower ResNet18 accuracy of $63.96\%$, increasing
to $70.31\%$ with their sequential multi-head method. The experiments
are not directly equivalent because the models and preprocessing differ,
but the contrast demonstrates why evaluation protocol must be considered
alongside headline accuracy.

\subsection{DeepShip Performance with Recording-Level Partitioning}

\begin{table}[h]
\centering
\setlength{\tabcolsep}{1.5pt}
\renewcommand{\arraystretch}{1.0}

\caption{Selected DeepShip classification results and published
comparisons.}
\label{tab:deepship_results}

\begin{tabular*}{\columnwidth}{
@{\extracolsep{\fill}}
p{0.27\columnwidth}
p{0.21\columnwidth}
ccc
@{}}
\toprule
\textbf{Feature} &
\textbf{Model} &
\textbf{Params} &
\textbf{Val. F1} &
\textbf{Test} \\
\midrule

Mel   & 4L-CNN & 157K & 0.6519 & \textbf{0.7226 F1} \\
STFT  & 4L-CNN & 157K & 0.6351 & 0.7129 F1 \\
MFCC  & 4L-CNN & 149K & 0.6162 & 0.6735 F1 \\
CQT   & 4L-CNN & 173K & 0.6408 & 0.6902 F1 \\
BM    & 4L-CNN & 154K & \textbf{0.6766} & 0.6694 F1 \\
IHC   & 4L-CNN & 154K & 0.6562 & 0.6336 F1 \\

\midrule

STFT &
\shortstack[l]{RBF-SVM\\($C=0.1$)}
& N/A & 0.6266 & 0.6871 F1 \\

Mel &
\shortstack[l]{LogReg\\($C=0.03$)}
& N/A & 0.6380 & 0.6827 F1 \\

MFCC &
\shortstack[l]{LogReg\\($C=1$)}
& N/A & 0.6376 & 0.6598 F1 \\

CQT &
\shortstack[l]{RBF-SVM\\($C=1$)}
& N/A & 0.5830 & 0.7027 F1 \\

\midrule

Log-Mel &
ShuffleFAC-16 \cite{park2026shufflefac}
& 39K & -- & 0.7145 F1 \\

\shortstack[l]{LOFAR\\(STFT, $0$--$3$ kHz)} &
Mobile\_ViT \cite{yao2024mobilevit}
& 21.61M & -- & 0.9457 Acc. \\

\shortstack[l]{ZCR, RMS\\MFCC,Chroma} &
DCMT \cite{mahmud2026dcmt}
& 0.70M & -- & 0.9753 Acc. \\

\bottomrule
\end{tabular*}

\vspace{1mm}

\noindent
Rows above the final divider are from this study; published results
below use different evaluation protocols. ``--'' denotes an unreported
metric.

\end{table}

DeepShip provides the primary test of vessel classification when source recordings are explicitly separated before segmentation. The final $5$~second manifest contains no source-recording overlap between the training, validation, and test partitions, so the scores in Table~\ref{tab:deepship_results} measure performance on segments derived from unseen recordings. For each conventional representation, the classical classifier shown in Table~\ref{tab:deepship_results} is the configuration selected by
validation macro F1 rather than by test performance. Additionally, the compact four-layer CNN achieves substantially more moderate performance than the ShipsEar diagnostic benchmark, with observed test macro-F1 scores ranging from $0.6336$ to $0.7226$
across the evaluated acoustic representations.

BM produces the highest validation macro F1, $0.6766$, whereas
Mel produces the highest observed test macro F1, $0.7226$. Because model
and feature selection must be based on validation rather than test
performance, the higher Mel test score is reported as an observed
held-out result rather than used retrospectively to select Mel as the
best representation. More broadly, the spread between the completed
feature results shows that representation choice can alter
recording-level performance substantially without requiring a larger
classifier.

The published results provide context rather than direct ranking
comparisons. ShuffleFAC-16 uses a recording-level $7{:}1{:}2$ split
before generating non-overlapping $3$~second log-Mel segments and reports
$0.7145$ macro F1 with $39$K parameters
\cite{park2026shufflefac}. Mobile\_ViT uses a LOFAR representation
derived from STFT over $0$--$3$~kHz and reports $94.57\%$ accuracy
\cite{yao2024mobilevit}, while DCMT combines ZCR, RMS, MFCC, and Chroma
features and reports $97.53\%$ accuracy \cite{mahmud2026dcmt}. Their
segmentation, partitioning, augmentation, and reported metrics differ
from the present experiment, so these values are not treated as
directly comparable to our recording-level macro F1 results.

To test whether the DeepShip generalisation gap was specific to one
split, we repeated the Mel RBF-SVM experiment using five-fold grouped
cross-validation, with all segments from each source recording confined
to a single fold. Mean validation macro F1 was $0.6200\pm0.0217$ at segment level and
$0.6974\pm0.0309$ after recording-level aggregation, with a mean
train--validation gap of $0.3196$. The consistent gap across folds
indicates that generalisation to unseen recordings remains difficult
rather than being specific to one validation partition.

\subsection{Grad-CAM Analysis}

\begin{figure}[h]
    \centering
    \includegraphics[width=\columnwidth]{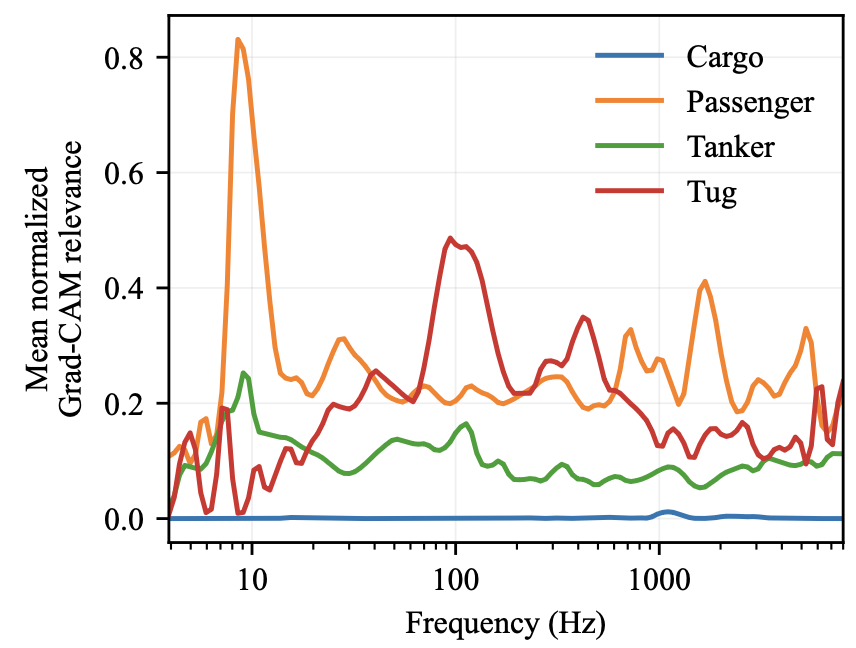}
    \caption{Class-wise mean normalized Grad-CAM relevance as a function of frequency for correct DeepShip test predictions using the four-layer STFT
CNN. The frequency profiles are obtained by averaging the Grad-CAM maps
over time and across predictions within each class.}
    \label{fig:gradcam_frequency}
\end{figure}

Figure~\ref{fig:gradcam_frequency} shows class-dependent frequency
relevance from Grad-CAM analysis of correct DeepShip test prediction results.
Passenger predictions concentrate strongly
at very low frequencies, while Tug places greater relevance in the
lower-to-mid frequency region. Tanker shows a broader relevance profile,
whereas Cargo is comparatively sparse across frequency. These differences
show that the compact CNN does not rely on the same spectral evidence for
each vessel class, providing additional evidence that the STFT
representation preserves class-dependent structure used by the model.

\subsection{Effect of Model Capacity on DeepShip}

Increasing model capacity did not improve recording-level
generalisation under the matched $5$~second DeepShip protocol. The
$11.17$M-parameter ResNet18 reached a best validation macro F1 of
$0.6110$, compared with $0.6519$ for the approximately
$157$K-parameter Mel CNN. Thus, a roughly $71\times$ increase in
parameter count provided no validation improvement, indicating that
greater model capacity alone does not resolve the generalisation
difficulty.

We also performed a best-effort reproduction of the published DeepShip
ResNet18 baseline in \cite{chen2026uatfsn}, following its reported
$3$~second segmentation and recording-level $80{:}20$ protocol. The
reconstruction recovered the same $609$ recordings and $56{,}468$
segments. By epoch 40, training accuracy reached $99.47\%$, while
held-out accuracy remained $62.13\%$ with a macro F1 of $0.6192$,
substantially below the reported $95.13\%$ accuracy. Because held-out
performance was monitored when training was stopped, this result is
treated as diagnostic rather than an unbiased test estimate. The
remaining discrepancy cannot be attributed to a single factor because
several implementation and split details are not specified in the
publication.

\section{Conclusion}
\label{sec:conclusion}
This study shows that evaluation protocol is critical when assessing
compact models for underwater vessel classification. On the available
ShipsEar split, a two-layer STFT CNN reaches $0.9918$ macro F1 and a
pooled STFT RBF-SVM reaches $0.9883$, demonstrating that strong
classification performance can be obtained without large models.
However, the absence of recoverable source-recording provenance prevents
these results from being interpreted as recording-independent
generalisation. When source recordings are explicitly separated on
DeepShip, performance becomes substantially more moderate, with the
compact four-layer CNN reaching observed test macro F1 scores of up to
$0.7226$. Under the matched $5$~second DeepShip protocol, the
$11.17$M-parameter ResNet18 also achieves lower validation macro F1 than
the approximately $157$K compact CNN, showing that increasing model
capacity alone does not improve recording-level generalisation.

The results instead support greater emphasis on acoustic representation
and evaluation design before increasing classifier capacity. Conventional
and auditory-inspired representations both provide useful
vessel-discriminative information, while statistical pooling enables
this information to be used by comparatively small classifiers. Grad-CAM
analysis further shows class-dependent spectral relevance within the
STFT CNN, providing additional evidence that the representation retains
discriminative structure used by the model. Together, these findings
support compact, representation-driven approaches as a practical
direction for deployable underwater acoustic classification.

\clearpage


\bibliographystyle{IEEEbib}
\bibliography{references}

@article{santosdominguez2016shipsear,
  author  = {Santos-Dom{\'i}nguez, David and Torres-Guijarro, Soledad and Cardenal-L{\'o}pez, Antonio and Pena-Gimenez, Antonio},
  title   = {{ShipsEar}: An Underwater Vessel Noise Database},
  journal = {Applied Acoustics},
  volume  = {113},
  pages   = {64--69},
  year    = {2016},
  doi     = {10.1016/j.apacoust.2016.06.008}
}

@article{irfan2021deepship,
  author  = {Irfan, Muhammad and Zheng, Jiangbin and Ali, Shahid and Iqbal, Muhammad and Masood, Zafar and Hamid, Umar},
  title   = {{DeepShip}: An Underwater Acoustic Benchmark Dataset and a Separable Convolution Based Autoencoder for Classification},
  journal = {Expert Systems with Applications},
  volume  = {183},
  pages   = {115270},
  year    = {2021},
  doi     = {10.1016/j.eswa.2021.115270}
}

@book{lyon2017human,
  author    = {Lyon, Richard F.},
  title     = {Human and Machine Hearing: Extracting Meaning from Sound},
  publisher = {Cambridge University Press},
  year      = {2017},
  doi       = {10.1017/9781139051699}
}

@misc{lyon2024carfacv2,
  author        = {Lyon, Richard F. and Schonberger, Rob and Slaney, Malcolm and Velimirovi{\'c}, Mihajlo and Yu, Honglin},
  title         = {The {CARFAC} v2 Cochlear Model in {Matlab}, {NumPy}, and {JAX}},
  year          = {2024},
  eprint        = {2404.17490},
  archivePrefix = {arXiv},
  primaryClass  = {eess.AS},
  doi           = {10.48550/arXiv.2404.17490}
}

@inproceedings{selvaraju2017gradcam,
  author    = {Selvaraju, Ramprasaath R. and Cogswell, Michael and Das, Abhishek and Vedantam, Ramakrishna and Parikh, Devi and Batra, Dhruv},
  title     = {{Grad-CAM}: Visual Explanations from Deep Networks via Gradient-Based Localization},
  booktitle = {Proceedings of the IEEE International Conference on Computer Vision (ICCV)},
  pages     = {618--626},
  year      = {2017},
  doi       = {10.1109/ICCV.2017.74}
}

@article{park2026shufflefac,
  author  = {Park, Sangwon and Kim, Dongjun and Byun, Sung-Hoon and Park, Sangwook},
  title   = {Ultra-Lightweight Ship-Radiated Sound Classification for Real-Time Embedded Inference},
  journal = {IEEE Embedded Systems Letters},
  year    = {2026},
  doi     = {10.1109/LES.2026.3708295}
}

@article{yao2024mobilevit,
  author  = {Yao, Haiyang and Gao, Tian and Wang, Yong and Wang, Haiyan and Chen, Xiao},
  title   = {{Mobile\_ViT}: Underwater Acoustic Target Recognition Method Based on Local--Global Feature Fusion},
  journal = {Journal of Marine Science and Engineering},
  volume  = {12},
  number  = {4},
  pages   = {589},
  year    = {2024},
  doi     = {10.3390/jmse12040589}
}

@article{mahmud2026dcmt,
  author  = {Mahmud, Nahid-Al and Zhang, Tao and Iqbal, Yasir and Sumona, Farhana Bari and Azaz, Ikram and Geng, Yanzhang and Khan, Wajid and Kharma, Qasem M. and Rubanenko, Oleksandr},
  title   = {An Efficient Transformer Architecture with Depthwise Separable Convolutions for High-Accuracy Underwater Acoustic Target Recognition},
  journal = {Scientific Reports},
  volume  = {16},
  pages   = {2733},
  year    = {2026},
  doi     = {10.1038/s41598-025-32401-3}
}

@article{yan2026multhead,
  author  = {Yan, Chenhong and Yan, Shefeng and Yu, Yang and Gao, Ruobin and Pan, Guang and Yang, Yankun and Yao, Tianyi and Suganthan, Ponnuthurai N.},
  title   = {Boosting-Inspired Sequential Multi-Head Learning for Underwater Acoustic Target Recognition},
  journal = {Ocean Engineering},
  volume  = {356},
  pages   = {125324},
  year    = {2026},
  doi     = {10.1016/j.oceaneng.2026.125324}
}

@article{brown1991cqt,
  author  = {Brown, Judith C.},
  title   = {Calculation of a Constant Q Spectral Transform},
  journal = {The Journal of the Acoustical Society of America},
  volume  = {89},
  number  = {1},
  pages   = {425--434},
  year    = {1991},
  doi     = {10.1121/1.400476}
}

@article{hohmann2002gammatone,
  author  = {Hohmann, Volker},
  title   = {Frequency Analysis and Synthesis Using a Gammatone Filterbank},
  journal = {Acta Acustica united with Acustica},
  volume  = {88},
  number  = {3},
  pages   = {433--442},
  year    = {2002}
}

@article{chen2026uatfsn,
  author  = {Chen, Menghan and Lu, Yuchen and Cheng, Liangliang and Zhu, Rongxin and Tao, Kun and Li, Yifei and Abdel Wahab, Magd},
  title   = {Lightweight Underwater Acoustic Time-Frequency Separation Network for Efficient Marine Target Recognition},
  journal = {Ocean Engineering},
  volume  = {343},
  pages   = {123234},
  year    = {2026},
  doi     = {10.1016/j.oceaneng.2025.123234}
}

@InProceedings{He_2016_CVPR,
author = {He, Kaiming and Zhang, Xiangyu and Ren, Shaoqing and Sun, Jian},
title = {Deep Residual Learning for Image Recognition},
booktitle = {Proceedings of the IEEE Conference on Computer Vision and Pattern Recognition (CVPR)},
month = {June},
year = {2016}
}

@article{toma2018smart,
  author  = {Toma, Daniel Mihai and Masmitja, Ivan and del R{\'i}o, Joaqu{\'i}n
             and Martinez, Enoc and Artero-Delgado, Carla and Casale, Alessandra
             and Figoli, Alberto and Pinzani, Diego and Cervantes, Pablo
             and Ruiz, Pablo and Mem{\`e}, Simone and Delory, Eric},
  title   = {Smart Embedded Passive Acoustic Devices for Real-Time Hydroacoustic Surveys},
  journal = {Measurement},
  volume  = {125},
  pages   = {592--605},
  year    = {2018},
  doi     = {10.1016/j.measurement.2018.05.030}
}

\end{document}